\documentclass[11pt,preprint,flushrt]{aastex}

\def\eqq#1{Equation~(\ref{#1})}

\newcommand{\bfr}{\mbox{\bf r}}

\begin{document}

\slugcomment{Submitted to ApJ}

\title{Multipole Formulae for Gravitational Lensing Shear and Flexion}

\author{Gary M. Bernstein\altaffilmark{1}}
\email{garyb@physics.upenn.edu}
\and
\author{Reiko Nakajima\altaffilmark{1,2}}
\email{rnakajima@berkeley.edu}
\altaffiltext{1}{Department of Physics \& Astronomy, University of Pennsylvania, 
209 S.\ 33rd St., Philadelphia, PA 19104}
\altaffiltext{2}{Lawrence Berkeley Laboratory, 1 Cyclotron Road,
  Berkeley, CA 94720}

\begin{abstract}
The gravitational lensing equations for convergence, potential, shear,
and flexion are simple in polar coordinates and separate under a
multipole expansion once the shear and flexion spinors are rotated
into a ``tangential'' basis.  We use this to investigate whether the
useful monopole aperture-mass shear formulae generalize to all
multipoles and to flexions. We re-derive the result of Schneider and
Bartelmann that the shear multipole $m$ at radius $R$ is completely
determined by the mass multipole at $R$, plus specific moments $Q^{(m)}_{\rm
  in}$ and $Q^{(m)}_{\rm out}$ of the mass multipoles internal and
external, respectively, to $R$. The $m\ge0$ multipoles are independent
of $Q_{\rm out}$. But in contrast to the monopole, the $m<0$ multipoles
are independent of $Q_{\rm in}$.  These internal and
external mass moments can be determined by shear (and/or flexion) data
on the complementary portion of the plane, which has practical
implications for lens modelling.  We find that the ease of $E/B$
separation in the monopole aperture moments does {\em not} generalize
to $m\ne 0$: the internal monopole moment is the {\em only} non-local
E/B discriminant available from lensing observations.  We have also
not found practical {\em local} E/B discriminants beyond the monopole,
though they could exist. 
We show also that the use of weak-lensing data to constrain a constant
shear term near a strong-lensing system is impractical without strong
prior constraints on the neighboring mass distribution.
\end{abstract}


\section{Introduction}
Weak gravitational lensing measurements of the shear $\gamma$ are
often used to constrain the mass distributions in galaxies, cluster of
galaxies, or even larger-scale objects.  An exceptionally useful set
of {\em aperture mass formulae} give non-parametric relations between
the monopole moments of the lensing shear and the lensing mass.  The
aperture-mass formulae have these interesting aspects:
\begin{enumerate}
\item
A relation between
the mean tangential shear component $\gamma_t$ on a
circle of radius $R$, and the mean convergence at and within $R$ \citep{K95}:
\begin{equation}
\label{apmass1}
\langle \gamma_t \rangle_R = \langle \kappa \rangle_{<R} - \langle
\kappa \rangle_R.
\end{equation}
Recall that the convergence $\kappa$ is the surface mass density in
units of the lensing critical density.  The monopole of the
tangent shear component at $R$ is dependent upon the mass at or
interior to $R$ in a simple way, and is independent of the mass
exterior to $R$.
\label{prop1}
\item
A rearrangement of the aperture-mass formula is \citep{Fahl94}
\begin{equation}
\label{apmass2}
\langle \kappa \rangle_{<R} = 2 \int_R^\infty dr \,r^{-1}\, \langle \gamma_t
\rangle_r.
\end{equation}
This relation allows a model-independent determination of the mass
monopole within $R$ using only shear measures at $r\ge R$.  A generalization to
radially weighted aperture masses is given by \citet{KSFW94} and
\citet{Sch96}, which can allow the shear integral to have finite
support.
\label{prop2}
\item
The shear monopole admits an
instantaneous test for the presence of ``B-mode'' deflections.  If we allow
the lensing potential to be sourced by the normal scalar mass
distribution $\kappa_E$ plus a pseudo-scalar mass $\kappa_B$, then we
find that the tangent-shear formulae (\ref{apmass1}) applies to only
the E-mode mass.  The B-mode mass produces a monopole of the ``skew'' shear
component $\gamma_s$ rotated 45\arcdeg\ from the tangent
direction.\footnote{The skew shear component is frequently designated
  by the misnomer ``radial shear.''}  The skew shear is a direct test
for B-mode sources within the aperture:
\begin{equation}
\label{apmass3}
\langle \gamma_s \rangle_R = \langle \kappa_B \rangle_{<R} - \langle
\kappa_B \rangle_R.
\end{equation}
Hence in real observations the monopole skew shear should be null.
This is a special case of the general rule that any E-mode mass
measurement should be nulled when all shears are rotated by
45\arcdeg\ \citep{SMF96, LK97}.
\label{prop3}
\end{enumerate}

In this paper we ask: can these three useful formulae relating
monopole shear moments to monopole mass moments be extended to
multipole moments? \citet{SB97}[SB97] offer a division of the shear
multipoles into internal and external terms (their Appendix B),
generalizing property (1).  They further extend property (2), the
ability to determine the mass moment from a closed-form integral of
the shear, to the general multipole case.  We offer here a simpler
re-derivation of their results, extending them to the case where
B-mode lensing may be present.  In the process we also inquire whether
property (3) can be extended: is
there a simple test for $B$-mode mass in the shear multipole signals?

We further ask: are there equivalent properties for the higher-order
lensing distortions, {\it i.e.} ``flexions'' \citep{BGRT}?  The
expected answer is yes, since these are sourced by the same two
scalar degrees of freedom $\kappa_{E,B}$ that produce the shear field.

In \S\ref{derive} we derive a simple differential relation between the
moments of convergence (mass), shear, and flexion, using a very compact
notation for the standard lensing equations.  This derivation will
allow for both E- and B-mode source terms.  In \S\ref{formulae} we
will examine the multipole generalizations of
Equations~(\ref{apmass1})--(\ref{apmass3}).  In \S\ref{app} we give an
application of these multipole formulae to a common problem in lens
modelling: a galaxy-scale strong lens is embedded in a more
extended group or cluster potential.  We show how weak shear
measurements could be used to constrain the cluster potential without
assuming a particular geometry for the cluster mass.

\section{Lensing formulae in polar coordinates}
\label{derive}
First we recast the familiar lensing equations into polar coordinates.
In the flat-sky limit, lensing is compactly described by the
differential operators of \citet{CHK05}
\begin{eqnarray}
\partial & \equiv & {\partial \over \partial x} + i{\partial \over \partial
  y} = e^{i\theta}\left( {\partial \over \partial r} + {i \over r}
 {\partial \over \partial\theta}\right) \\
\bar\partial & \equiv & {\partial \over \partial x} - i{\partial \over \partial
  y} = e^{-i\theta}\left( {\partial \over \partial r} - {i \over r}
 {\partial \over \partial\theta}\right).
\end{eqnarray}
For an arbitrary deflection field $(\alpha_x, \alpha_y)$ defined on
the plane of the sky, we define a complex deflection
$\alpha\equiv\alpha_x + i \alpha_y$.  The deflection field can be
decomposed into a curl-free $E$-mode part and divergence-free $B$-mode
part by defining a complex potential $\psi=\psi_E + i \psi_B$ from the
scalar potential $\psi_E$ and pseudoscalar potential $\psi_B$:
\begin{equation}
\alpha = \partial \psi.
\end{equation}
The shear imposed on the background sources is given by the
derivatives of the deflection:
\begin{equation}
\gamma \equiv \gamma_1 + i \gamma_2 = {1 \over 2} \partial\alpha.
\end{equation}
Furthermore the complex convergence $\kappa \equiv \kappa_E + i
\kappa_B$, which is the source term for the complex potential, can be
written as $2\kappa = \bar\partial\alpha$.
Then the convergence and shear can be expressed in terms of the
projected potential $\psi$ as
\begin{eqnarray}
2\kappa & = & \nabla^2\psi = \partial \bar \partial \psi \\
2\gamma & = & \partial \partial \psi.
\label{shearpotential}
\end{eqnarray}
From this we immediately derive the \citep{K95} relation between
convergence and shear, which holds even when $B$-modes are present:
\begin{equation}
\partial\kappa = \bar\partial\gamma.
\label{kaiser}
\end{equation}
The shear components defined with respect to the radius vector can be
expressed as 
\begin{equation}
\Gamma \equiv \gamma_t + i\gamma_s = -\gamma e^{-2i\theta}.
\end{equation}
Substituting this into the Kaiser relation (\ref{kaiser}) yields
\begin{equation}
\kappa_{,r} + {i \over r}\kappa_{,\theta}
= -\Gamma_{,r} + {i \over r} \Gamma_{,\theta} - {2 \over r}\Gamma
\end{equation}
where the subscripts after the comma denote differentiation, as
usual. From \eqq{shearpotential}, the shear can also be expressed as
\begin{equation}
-2\Gamma = \left[ {\partial^2 \over \partial r^2}
 -{1 \over r}{\partial \over \partial r}
 -{1 \over r^2} {\partial^2 \over \partial \theta^2}
 + 2i {\partial \over \partial r}
\left( {1 \over r} {\partial \over \partial\theta}\right) \right]
\psi.
\end{equation}
Both of these equations clearly separate into radial and azimuthal
parts under a multipole decomposition of the relevant quantities.
For any complex quantity $z$ we define the multipoles as
\begin{eqnarray}
z(\bfr) & = & {1 \over 2} \sum_{m=-\infty}^\infty (1+\delta_m)z^{(m)}(r) e^{im\theta} \\
z^{(m)}(r) & = & {1 \over (1+\delta_m)\pi} \int d\theta
z(\bfr)e^{-im\theta}.
\end{eqnarray}
The Kaiser relation can be rewritten for each multipole as
\begin{eqnarray}
\kappa^{(m)}_{,r} - {m \over r}\kappa^{(m)}
& = & -\Gamma^{(m)}_{,r} - {m+2 \over r}\Gamma^{(m)} \\
= r^m{\partial \over \partial r}(r^{-m}\kappa^{(m)})
 & = & -r^{-m-2}{\partial \over \partial r}(r^{m+2}\Gamma^{(m)}).
\label{kG1}
\end{eqnarray}
\subsection{Flexions}
The third derivatives of the lensing potential can be described by two
complex-valued ``flexion'' fields \citep{CHK05,BGRT}: the spin-1 field
${\cal F}\equiv \partial \partial \bar\partial \phi/2$, and the spin-3
field ${\cal G} \equiv \partial \partial \partial \phi/2$.  These
satisfy 
\begin{equation}
\label{flex1}
{\cal F} = \partial \kappa, \qquad 
\bar\partial{\cal G} = \partial\partial\kappa.
\end{equation}
In analogy with the shear, we define
flexions in a tangential basis:
\begin{equation}
F \equiv e^{-i\theta}{\cal F}, \qquad G \equiv e^{-3i\theta}{\cal G}.
\end{equation}
Casting \eqq{flex1} into polar coordinates, performing a
multipole decomposition of $F$ and $G$, then defining $H\equiv G-F$,
we obtain formulae for flexion multipoles in terms of convergence:
\begin{eqnarray}
\label{flex2}
r^m{\partial \over \partial r}(r^{-m}\kappa^{(m)})
 & = & F^{(m)} \\ \nonumber
 & = & -r^{-m-2} {\partial \over \partial r} \left(
{r^{m+3} H^{(m)} \over 2(m+2)} \right).
\end{eqnarray}
Note the $\kappa$ dependence on the left-hand side is identical to
\eqq{kG1}; not surpringly, the flexion multipoles are very close to
the shear multipoles.  In particular
\begin{equation}
H^{(m)} = {2(m+2) \over r}\Gamma^{(m)}
\label{flex3}
\end{equation}

\section{Multipole formulae}
\label{formulae}
\subsection{Interior and Exterior Shears}
Several useful results may be obtained by integrating \eqq{kG1} by
parts.  First, we obtain a closed-form expression for the shear
multipoles: 
\begin{equation}
\left . r^{m+2} \kappa^{(m)} \right|^{R_2}_{R_1}
 - 2(m+1) \int_{R_1}^{R_2}dr\,r^{m+1}\kappa^{(m)}(r)
= -\left . r^{m+2} \Gamma^{(m)} \right|^{R_2}_{R_1}
\label{r1r2}
\end{equation}
For $m\ge0$ we can assume $R_1^{m+2} \kappa^{(m)}(R_1)\rightarrow
0$ and $R_1^{m+2} \Gamma^{(m)}(R_1)\rightarrow
0$ as $R_1\rightarrow 0$ for any mass distribution which remains
finite and differentiable at the origin.
In this case
\begin{equation}
\Gamma^{(m)}(R) = -\kappa^{(m)}(R) + {2(m+1) \over R^{m+2}}
\int_0^R r\,dr\, r^m\kappa^{(m)}(r) \qquad (m\ge0).
\label{gam}
\end{equation}
Since $\Gamma^{(m)}$
is completely determined
by the mass distribution at and interior to $R$ for $m\ge0$,
this is the desired
generalization monopole formula (\ref{apmass1}).
$\Gamma(r)$ is, however,
a complex quantity, so it is not fully specified by the $m\ge0$
multipoles.  For $m<0$ a bounded mass distribution will have 
$R_2^{m+2} \kappa^{(m)}(R_2)\rightarrow
0$ and $R_2^{m+2} \Gamma^{(m)}(R_2)\rightarrow
0$ as $R_2\rightarrow\infty$.  We then obtain
\begin{equation}
\Gamma^{(m)}(R) = -\kappa^{(m)}(R) - {2(m+1) \over R^{m+2}}
\int_R^\infty r\,dr\, r^m\kappa^{(m)}(r) \qquad (m<0).
\label{gamneg}
\end{equation}
The negative-$m$ multipoles of $\Gamma$ are hence dependent only on mass at or
{\em exterior} to $R$.  This
formula and the previous one fully specify the
shear field.  So the shear in a region $R_1<r<R_2$ is
determined completely by the mass distribution in this region, plus
these multipole moments of the mass interior and exterior to the
annulus:
\begin{eqnarray}
Q^{(m)}_{\rm in}(R) & \equiv & \int_{r<R} d^2r\, r^m e^{-im\theta} \kappa(\bfr) \\
 & = & (1+\delta_m)\pi\int_0^R r\,dr\,r^m \kappa^{(m)}(r) \\
\label{moments}
 & = & \pi R^2 \langle\kappa\rangle_{<R} \left\langle (x-iy)^m \right\rangle, \\
Q^{(m)}_{\rm out}(R) & \equiv & \int_{r>R} d^2r\, r^{-m} e^{im\theta} \kappa(\bfr) \\
 & = & (1+\delta_m)\pi\int_R^\infty r\,dr\,r^{-m}\kappa^{(-m)}(r).
\end{eqnarray}
These definitions are normalized to agree with
Eqns~(B5) of SB97, who similarly demonstrate that shear at $R$
depends upon interior and exterior masses only through these
quantities. We have altered the phase conventions, however, in order
to work successfully with the complex (E and B) convergence and
potentials. 
The brackets in 
\eqq{moments} indicate a mass-weighted average inside radius $R$.

The present derivation shows that
the division into interior and exterior shears holds even
when there is an imaginary ($B$-mode) component to the potential and
convergence.  The shear components $\gamma_t,$ and
$\gamma_s$ are always real-valued, as is $\kappa$ when there is no $B$-mode
lensing: $\bar\kappa^{(-m)}=\kappa^{(m)},$ etc. But this relation need
not hold for $\Gamma^{(m)}$.  It remains true, however, that the $m\ge0$
multipoles are produced by mass internal to $R$ while $m<0$ are
produced by external mass.

Eqns~(\ref{gam}) and (\ref{gamneg}) can be
restated as
\begin{eqnarray}
\Gamma^{(m)}(r) 
 & = & -\kappa^{(m)}(r) +
{2 (m+1) \over (1+\delta_m) \pi r^{m+2}} Q^{(m)}_{\rm
    in}(r) \qquad (m\ge0) \\
\Gamma^{(m)}(r)
 & = & -\kappa^{(m)}(r) -  
{2 (m+1) \over \pi r^{m+2}} Q^{(-m)}_{\rm
    out}(r) \qquad (m<0) 
\label{qout}
\end{eqnarray}

For the flexions, \eqq{flex3} makes it clear that $H^{(m)}$ depends on
local, internal, and external mass exactly as $\Gamma^{(m)}$ does.
\eqq{flex2} shows that the $F$ flexion depends only upon the local
value of $\kappa^{(m)}$ and its first derivative, independent of both
the internal and external moments $Q^{(m)}$.

\subsection{Mass multipoles from Shear}
Multiplying \eqq{kG1} by $r^{-m}$, integrating by parts, and taking
the upper integration limit to infinity yields
\begin{equation}
\kappa^{(m)}(R) = -\Gamma^{(m)}(R) +2(m+1)R^m \int_R^\infty
r\,dr\,\Gamma^{(m)}(r) r^{-m-2} \qquad (m\ge0).
\end{equation}
Comparison to \eqq{gam} yields
\begin{equation}
\label{qinsolve}
{ Q_{\rm in}^{(m)}(R) \over (1+\delta_m) \pi } = 
\int_0^R r\,dr\,r^m\kappa^{(m)}(r) = 
R^{2m+2} \int_R^\infty r\,dr\,r^{-m-2} \Gamma^{(m)}(r) \qquad (m\ge0).
\end{equation}
This is the desired generalization of the monopole aperture-mass
formula (\ref{apmass2}).  Multiplication by a weight function before
the integration by parts would yield the full weighted
aperture-multipole formulae of SB97.
The mass interior to $R$ affects the shear exterior
to $R$ only through the multipole moments $Q^{(m)}_{\rm in}(R)$.  Here
we see that these moments are completely recoverable from the shear
field $\Gamma^{(m)}$ exterior to $R$.  The $Q^{(m)}_{\rm in}$ are thus
a complete description of the information that lensing data exterior
to $R$ can offer on the mass distribution interior to $R$.  This holds
even in the presence of B-mode lensing.

Taking $m<0$ in this integration by parts yields an analogous formula
by which the $Q^{(m)}_{\rm out}(R)$ values may be determined from
shear data at $r<R$:
\begin{equation}
{ Q_{\rm out}^{(-m)}(R) \over \pi } = 
\int_R^\infty r\,dr\,r^m\kappa^{(m)}(r) = 
R^{2m+2} \int_0^R r\,dr\,r^{-m-2} \Gamma^{(m)}(r). \qquad (m<0).
\end{equation}

When the shear data is available in a finite annulus $R_1<r<R_2$, we
can take differences of the above two formulae:
\begin{eqnarray}
\label{annulus1}
\int_{R_1<r<R_2}d^2r \left[\gamma_t(\bfr)+i\gamma_s(\bfr)\right]
r^{-m-2}e^{-im\theta} & = & 
R_1^{-2m-2} Q^{(m)}_{\rm in}(R_1) - R_2^{-2m-2} Q^{(m)}_{\rm
  in}(R_2) \qquad (m\ge 0) \\
\label{annulus2}
\int_{R_1<r<R_2}d^2r \left[\gamma_t(\bfr)+i\gamma_s(\bfr)\right]
r^{m-2}e^{im\theta} & = & 
- R_1^{2m-2} Q^{(m)}_{\rm out}(R_1) + R_2^{2m-2} Q^{(m)}_{\rm
  out}(R_2) \qquad (m\ge 1).
\end{eqnarray}

\subsection{Mass multipoles from Flexion}
\eqq{flex3} implies that mass multipoles can be retrieved from the
$H^{(m)}$ using the preceding shear formulae.  Mass reconstruction
from the $F$ flexion differs: \eqq{flex2} integrates by parts to yield
\begin{equation}
\kappa^{(m)}(R) = \left\{
\begin{array}{ll}
-\int_R^{\infty} (r/R)^{-m} F^{(m)}(r)\,dr & m\ge0 \\
\int_0^R (r/R)^{-m} F^{(m)}(r)\,dr & m<0
\end{array}
\right.
\end{equation}

\subsection{E-B decomposition}
The real part of the monopole shear $\Gamma^{(0)}(R)$ at $R$ determines
the enclosed E-mode mass while the imaginary part gives the enclosed
B-mode mass.  This result does {\em not} generalize to other
multipoles.  Consider first the case where $\kappa^{(m)}(R)=0$, {\it
  i.e.} we are in a mass-free zone.  Then for $m\ge0$,
$\Gamma^{(m)}(R)$ is fully specified by the complex number
$Q^{(m)}_{\rm in}(R)$.  But for $m>0$, any chosen $Q^{(m)}_{\rm in}$
amplitude and phase produced by an E-mode source $\kappa_E$ can also
be produced by a pseudo-mass source $\kappa_B$ that is just the
$\kappa_E$ rotated about the origin by $(90/m)\arcdeg$.  There is
hence {\em no way to distinguish an internal $m>0$ E-mode mass
  distribution from an internal B-mode mass distribution.}  Similarly
we can produce any desired $Q^{(m)}_{\rm out}(R)$ with either
$\kappa_E$ or $\kappa_B$ source terms, so {\em there is no test that
  can distinguish E-mode mass from B-mode mass distributions external
  to the shear measurement zone.}  These conclusions hold for flexion
data as well as for shear data.

The monopole turns out to be a special case: the E/B diagnosis is
possible because the monopole 
E-mode and B-mode moments each have only 1 degree of freedom while the
observable $Q^{(0)}_{\rm in}$ is complex.  But for $m\ne 0$, the
E and B mass moments each have 2 degrees of freedom, so cannot be
independently retrieved from a single $Q$.

Thus if we have shear data on the $R_1<r<R_2$ annulus, we have hope
only of testing $\kappa$ for E and B components at $m\ne0$ only within
the annulus, 
not interior or exterior to it.  Ideally this can be done by noting
\begin{eqnarray}
2\kappa_E^{(m)} & = & \kappa^{(m)} + \bar \kappa^{(-m)} \\
2i\kappa_B^{(m)} & = & \kappa^{(m)} - \bar \kappa^{(-m)}.
\end{eqnarray}
This can be combined with \eqq{flex2}, for example, to yield a pure-E
quantity:
\begin{equation}
2\left( r\kappa^{(m)}_{E,rr} + \kappa^{(m)}_{E,rr} +
  m^2\kappa^{(m)}_E/r \right)
= r^{-m} {\partial \over \partial r} \left(r^{m+1} F^{(m)} \right)
+ r^m {\partial \over \partial r} \left(r^{1-m} \bar F^{(-m)}
\right).
\end{equation}
Sending $F\rightarrow iF$ gives a pure-B quantity.  These equations
are not practical null tests for B-modes, however, because they
involve derivatives of $F$, which have divergent noise in the presence
of shot noise from finite sampling.  A practical null test for $E/B$
modes in an annular region would require to an integral of $F$ (or
$\Gamma$ or $H$) over the annulus which could be approximated by a sum
over source galaxies.  We have not been able to derive such a form.

\section{Application to strong-lensing models}
\label{app}
In modeling a lensing system around a galaxy, one has strong-lensing
constraints from multiply-imaged sources.  The lens-mass model often
contains a galaxy mass distribution $\kappa_g(\bfr)$, but it is
essential in most cases to consider the influence of the larger-scale
mass distribution on the system.  Call this the ``cluster'' mass,
which generates potential $\psi_c(\bfr)$.  On the assumption that the
cluster mass has little structure on the scale of the strong-lensing
system, $\psi_c(\bfr)$ can be approximated by a few terms of a Taylor
expansion about the galaxy center within some radius $R_1$ that contains
all of the strongly lensed features \citep{Ko91}.  The constant and
linear terms of 
the Taylor expansion are immaterial to the strong-lensing model.
The potential at $r<R_1$, to cubic order in
the Taylor expansion of the cluster, is
\begin{equation}
\psi(\bfr) = (1-\kappa_c)\left[
  \psi_g(\bfr) + {\rm Re}\left( {\gamma \over 2}r^2 e^{-2i\theta}
    + {\sigma \over 4} r^3 e^{-i\theta} 
    + {\delta \over 6} r^3 e^{-3i\theta} \right) \right]
+ {\kappa_c \over 2} r^2,
\label{phiclust}
\end{equation}
with $\nabla^2 \psi_g = 2\kappa_g$.  The strong-lens data produces a
likelihood distribution over the (complex) parameters $\{\gamma,
\sigma, \delta\}$ and the galaxy-mass parameters.  The mass-sheet
degeneracy leaves $\kappa_c$ unconstrained by strong-lensing data.

We now ask what additional constraints on these model parameters are
available from the shear field at 
$r>R_1$.  We do {\em not} want to assume that the Taylor expansion
offers an adequate description of the cluster mass at $r>R_1$, but from
the previous discussion we know that only the multipole moments
$Q^{(m)}_{\rm out}(R_1)$ affect the parameters in the strong-lensing
potential.  In particular the terms of the form $r^me^{\pm im\phi}$ in
the Taylor expansion of $\phi_c$ can only be generated by mass {\em
  outside} $R_1$ 
while the other terms can only be generated by mass {\em inside}
$R_1$. Specifically:
\begin{itemize}
\item The $\kappa_c$ term is a monopole (constant) mass distribution, and 
$Q^{(0)}_{{\rm in},c}= \pi R_1^2\kappa_c$.
\item The $\gamma$ term is a constant shear, producing
  $\Gamma^{(-2)}=-2\gamma$. From \eqq{qout}, $Q^{(2)}_{{\rm
      out},c}(R_1)=-\pi\gamma$.
\item The $\sigma$ term is a dipole mass distribution, $\kappa={\rm
    Re}(\sigma r e^{-i\theta})$.  This being the only dipole cluster
  mass kept in the expansion, we have
  $Q^{(1)}_{{\rm in},c}= \pi R_1^4\bar\sigma/4$.
\item The $\delta$ term is an $m=3$ external shear,
  $\Gamma^{(-3)}=-2\delta r$, implying $Q^{(3)}_{{\rm
      out},c}(R_1)=-\pi\delta/2$.
\end{itemize}

Eqns~(\ref{annulus1}) and (\ref{annulus2}) can now be applied to a
shear measurement that extends to radius $R_2$ from the galaxy
center.  The multipole moments $Q$ are split into galaxy and cluster
contributions.  Those from the galaxy are calculable from the
parametric form adopted for $\kappa_g$.  The cluster contributions at
$R_1$ are parameterized by the Taylor expansion coefficients as
above.  The cluster contributions at $R_2$ are formally unconstrained,
but if $R_2$ is large enough then these may be bounded by even a rough
estimate of the total mass and extent of the cluster.  We obtain:
\begin{eqnarray}
\label{ann1}
\int_{R_1<r<R_2}d^2r \gamma_t(\bfr)
r^{-2} & = & \pi \kappa_c - R_2^{-2} Q^{(0)}_{{\rm in},c}(R_2) \\
\nonumber
 & & + (1-\kappa_c)\left[R_1^{-2}Q^{(0)}_{{\rm in},g}(R_1)
- R_2^{-2}Q^{(0)}_{{\rm in},g}(R_2)\right] \\
\label{ann2}
\int_{R_1<r<R_2}d^2r \left[\gamma_t(\bfr)+i\gamma_s(\bfr)\right]
r^{-3}e^{-i\theta} & = & \pi(1-\kappa_c)\bar\sigma/4 - R_2^{-4} Q^{(1)}_{{\rm in},c}(R_2) \\
\nonumber
 & & + (1-\kappa_c)\left[R_1^{-4}Q^{(1)}_{{\rm in},g}(R_1)
- R_2^{-4}Q^{(1)}_{{\rm in},g}(R_2)\right] \\
\label{ann3}
\int_{R_1<r<R_2}d^2r \left[\gamma_t(\bfr)+i\gamma_s(\bfr)\right]
e^{2i\theta} & = & \pi R_1^2(1-\kappa_c)\gamma + R_2^2 Q^{(2)}_{{\rm out},c}(R_2) \\
\nonumber
 & & + (1-\kappa_c)\left[-R_1^2Q^{(2)}_{{\rm out},g}(R_1)
+ R_2^2Q^{(2)}_{{\rm out},g}(R_2)\right] \\
\label{ann4}
\int_{R_1<r<R_2}d^2r \left[\gamma_t(\bfr)+i\gamma_s(\bfr)\right]
r e^{3i\theta} & = & \pi R_1^{4}(1-\kappa_c)\delta/2 +
R_2^{4} Q^{(3)}_{{\rm out},c}(R_2) \\ 
\nonumber
 & & + (1-\kappa_c)\left[-R_1^{4}Q^{(3)}_{{\rm out},g}(R_1)
+ R_2^{4}Q^{(3)}_{{\rm out},g}(R_2)\right].
\end{eqnarray}
Note that if the galaxy mass distribution has inversion symmetry about
the coordinate origin, then
all of the $Q^{(1)}$ and $Q^{(3)}$ moments of the galaxy vanish.
In each equation, the left-hand side is an observable quantity and the
right-hand side is a function of the galaxy parameters, the
Taylor-expansion parameters, and some (presumably small) terms for the
multipole moments to radius $R_2$.  Given {\it a priori} estimates of
the $R_2$ moments, each observation yields an additional constraint on
the strong-lens model, with the mass-sheet degeneracy now broken.

Were we to extend the cluster potential Taylor series to quartic
terms, we would have 5 new degrees of freedom.  The shear integrals
for $Q^{(2)}_{\rm in}$ and $Q^{(4)}_{\rm out}$ would constrain four of
these.  The $\psi \propto r^4$ term, however, would be an internal
monopole mass distribution with $r^2$ radial dependence.  It would
contribute to $Q^{(0)}_{\rm in}$ and would be degenerate with
$\kappa_c$ within the shear annulus.  The degeneracy would have to
be broken by the strong-lensing data.

The shear data on the left-hand sides are affected by the intrinsic
shape variation of the source galaxies, hence the constraints provided
by each equation are not exact.  As $R_2\rightarrow \infty$, the
negative powers of $r$ in the integrands of Equations~(\ref{ann1}) and
(\ref{ann2}) lead to bounded shape-noise uncertainties on
$\kappa_c$ and $\sigma$.  Unfortunately the shape noise in the
integrands of Equations~(\ref{ann3}) and
(\ref{ann4}) is divergent as the annulus grows outwards.  Hence these
expressions cannot be expected to provide useful constraints on the
cluster quadrupole and octupole mass moments $\gamma$ and
$\delta$. 

\subsection{(Constant) external shear}
Common practice in analysis of galaxy-scale strong-lensing systems is
to limit modelling of external mass to a constant shear across the
strong-lensing system, {\it i.e.} the $\gamma$ term of
\eqq{phiclust}. If we simplify \eqq{ann3} by setting $\kappa_c=0$ and
ignoring the shear induced by the galaxy mass, we find that the
observable quantity on the left-hand side is just the mean shear
inside the annulus, and
\begin{equation}
\gamma = {R_2^2 \over \pi R_1^2} Q^{(2)}_{\rm out}(R_2)
-{R_2^2-R_1^2 \over R_1^2} \langle \gamma \rangle_{\rm ann}.
\end{equation}
One approach is to use {\it a priori} knowledge of the mass
fluctuation spectrum to find a radius $R_2$ beyond which we can expect
the $Q^{(2)}_{\rm out}$ term to become negligibly small, $\lesssim
0.01$.  Note that in this case the desired shear $\gamma$ is {\em
  opposite} to the mean shear in the annulus.
Two problems arise however: first, it is not clear that any
such radius exists, since the large-scale ``cosmic shear'' is
typically $\sim0.01$ even before amplification by the $R_2^2/R_1^2$
factor in this term.  Second, the shape-noise variance of the measured 
$(R_2^2-R_1^2)\langle \gamma \rangle_{\rm ann}/R_1^2 $ contribution
will grow as $(R_2/R_1)^2$ for fixed $R_1$, rendering the measurement
uninteresting. It thus appears problematic to use weak-lensing
information to infer the ``external shear'' in galaxy-scale lenses.

In a different limit, one might assume that the mass {\em within} the
$R_1<r<R_2$ annulus has negligible quadrupole moment, perhaps because
one does not see any galaxy groups or clusters projected within this
annulus.  In this case, $Q^{(2)}_{\rm out}(R_2)=Q^{(2)}_{\rm
  out}(R_1)=-\pi\gamma$, and our estimate of the shear parameter
becomes simply {\em equal} to the mean shear in the annulus
\begin{equation}
\gamma = \langle \gamma \rangle_{\rm ann}.
\end{equation}
In this case the shape noise on $\gamma$ {\em decreases} as $R_2$ is
increased and can become usefully small.  

It thus appears that the use of weak-lensing information to constrain
the external shear on galaxy lenses is practical only if one has {\em
  a priori} knowledge that an annulus in the vicinity of the lens is
free of mass that would generate a significant shear on the system.
We note that these problems are exacerbated for the higher-order
external moments because of the positive powers of radius that appear
within the shear integrals, {\it e.g.} \eqq{ann4}.

\section{Conclusions}
Polar-coordinate expressions for the relations between convergence,
shear, and flexion are seperable under multipole expansions once we
rotate the shear and flexion spinors into ``tangential'' bases.  Two
well-known monopole aperture-mass properties are extensible to all
$m\ge0$ multipoles: first, the shear multipole $\Gamma^{(m)}(R)$ is
determined solely by the convergence (mass) multipole $\kappa^{(m)}$
at or interior to radius $R$.  The effect of the interior mass is
fully described by the moments $Q^{(m)}_{\rm in}(R) \propto \int_0^R
d^2r\,r^m \kappa^{(m)}$.  Second, we find that the value of the
interior mass moment $Q^{(m)}_{\rm in}(R)$ can be exactly recovered by
an integral of the shear multipole $\Gamma^{(m)}$ from $R$ to
$\infty$.

The multipoles $m<0$, however, behave oppositely to the monopole: the
shear at $R$ is determined exclusively by the mass at or {\em
  exterior} to $R$.  And the relevant exterior mass moment
$Q^{(m)}_{\rm out}(R)$ can be determined by
an integral of the shear {\em interior} to $R$.

The tangential flexion component $H\equiv e^{-3i\theta}{\cal G} -
e^{-i\theta}{\cal F}$ behaves exactly as the tangetial shear
$\Gamma$.  In fact they differ only by a factor $2(m+2)/r$.  The
vector flexion component $F$ depends purely on the local behavor of
$\kappa$, as is well known.

The simple $E/B$ decomposition of the monopole mass distribution does
{\em not} generalize to $m\ne0$.  We show that shear or flexion data
in an annulus $R_1<r<R_2$ cannot discriminate between E-mode and
B-mode sources outside this region---except for the monopole
case. Shear or flexion data may be able to distinguish $E$ from $B$
sources inside the annulus, but we have not been able to derive a
practical estimator which does this.

The multipole formulae presented by SB97 and extended here find
application in using weak-lensing data to constrain the large-scale
characteristics of mass distributions in the vicinity of
strong-lensing systems.  Unfortunately a complete characterization of
the exterior mass distributions is not practical because some of the aperture
multipole formulae have divergent shape-noise behavior.  In particular
the estimation of the constant ``external shear'' term often found in
strong-lens models is problematic without strong prior constraints on
the neighboring mass distributions.
We can
expect the aperture-multipole formulae to find further use in generating
model-independent measures of the shapes of dark-matter halos.

\acknowledgements
This work is
supported by grants AST-0607667 from the 
National Science Foundation, Department of Energy grant
DOE-DE-FG02-95ER40893 and NASA BEFS-04-0014-0018.

\end{document}